\documentclass[12pt,preprint]{emulateapj}

\usepackage{color}
\usepackage{amsmath}

\slugcomment{accepted for publication in the Astrophysical Journal Letters}

\shorttitle{Formation of Al$_2$O$_3$ Grains in Supernovae}
\shortauthors{Nozawa et al.}

\begin{document}

\title{PROBING THE PHYSICAL CONDITIONS OF SUPERNOVA EJECTA WITH
\\ THE MEASURED SIZES OF PRESOLAR A\lowercase{l}$_2$O$_3$ GRAINS}

\author{
Takaya Nozawa\altaffilmark{1},
Shigeru Wakita\altaffilmark{2},
Yasuhiro Hasegawa\altaffilmark{1,3}, and
Takashi Kozasa\altaffilmark{4} \\
}

\altaffiltext{1}{
Division of Theoretical Astronomy, National Astronomical Observatory of Japan, 
Mitaka, Tokyo 181-8588, Japan; takaya.nozawa@nao.ac.jp}
\altaffiltext{2}{Center for Computational Astrophysics, 
National Astronomical Observatory of Japan, Mitaka, Tokyo 181-8588, Japan}
\altaffiltext{3}{EACOA Fellow}
\altaffiltext{4}{Department of Cosmosciences, Graduate School of Science, 
Hokkaido University, Sapporo 060-0810, Japan}

%%%%%%%%%%%%%%%%%%%%%%%%%%%%%%%%%%%%%%%%%%%%%%%%%%%%%%%%%%%%%%%%%%%%%
\begin{abstract}

A few particles of presolar Al$_2$O$_3$ grains with sizes above 
0.5 $\mu$m are believed to have been produced in the ejecta of 
core-collapse supernovae (SNe).
In order to clarify the formation condition of such large Al$_2$O$_3$ 
grains, we investigate the condensation of Al$_2$O$_3$ grains for 
wide ranges of the gas density and cooling rate.
We first show that the average radius and condensation efficiency of
newly formed Al$_2$O$_3$ grains are successfully described by a 
non-dimensional quantity $\Lambda_{\rm on}$ defined as the ratio 
of the timescale with which the supersaturation ratio increases to the
collision timescale of reactant gas species at dust formation.
Then, we find that the formation of submicron-sized Al$_2$O$_3$ 
grains requires at least ten times higher gas densities than those 
presented by one-dimensional SN models.
This indicates that presolar Al$_2$O$_3$ grains identified as a SN 
origin might be formed in dense gas clumps, allowing us to propose 
that the measured sizes of presolar grains can be a powerful tool to 
constrain the physical conditions in which they formed.
We also briefly discuss the survival of newly formed Al$_2$O$_3$ 
grains against the destruction in the shocked gas within the SN
remnants.

\end{abstract}
%%%%%%%%%%%%%%%%%%%%%%%%%%%%%%%%%%%%%%%%%%%%%%%%%%%%%%%%%%%%%%%%%%%%%


%%%%%%%%%%%%%%%%%%%%%%%%%%%%%%%%%%%%%%%%%%%%%%%%%%%%%%%%%%%%%%%%%%%%%
\section{Introduction}
%%%%%%%%%%%%%%%%%%%%%%%%%%%%%%%%%%%%%%%%%%%%%%%%%%%%%%%%%%%%%%%%%%%%%

The role of core-collapse supernovae (SNe) as dust producers is a 
fundamental issue for understanding the evolution history of dust in 
the universe.
Recent far-infrared observations of young supernova remnants (SNRs) 
revealed that 0.1--1 $M_{\odot}$ of dust had formed in the metal-rich 
inner ejecta
(Sibthorpe et al.\ 2010; Barlow et al.\ 2010; Gomez et al.\ 2012; 
Matsuura et al.\ 2011, 2015; Indebetouw et al.\ 2014).
However, it remains to be clarified what fraction of these newly formed 
grains can survive the destruction in the shock-heated gas within the 
SNRs and be injected into the interstellar medium (ISM).
The destruction efficiency of dust grains depends on their chemical 
compositions and size distributions (e.g., Nozawa et al.\ 2007),  which 
are determined by the density and temperature evolution of the gas out 
of which they form (Nozawa \& Kozasa 2013) as well as the degree 
of mixing and clumpiness of the ejecta 
(Nozawa et al.\ 2003; Sarangi \& Cherchneff 2015).
Therefore, the physical condition and structure of the SN ejecta must be
fully appreciated through various approaches to unravel the properties of 
dust that is formed in the ejecta and is finally ejected to the ISM.

Presolar grains, which are identified in meteorites due to their highly 
anomalous isotopic compositions, are invaluable fossils that enable us 
to directly observe the detailed chemical compositions and sizes of stellar 
dust (Clayton \& Nittler 2004 and references therein).
Their isotopic signatures give clues about nucleosynthesis in stars and 
mixing of elements in the SN ejecta.
In addition, the measured sizes of presolar grains could offer key 
information on the physical conditions at their formation sites.

Among the presolar grains which are considered to have originated in SNe,
Al$_2$O$_3$ grains are of great importance because Al$_2$O$_3$ is 
believed to be one of the major dust components in Cassiopeia A (Cas A) 
SNR (Douvion et al.\ 2001; Rho et al.\ 2008).
Furthermore, most of the dust formation calculations have predicted the 
formation of Al$_2$O$_3$ grains in the ejecta of SNe as the first condensate
among oxide grains
(Kozasa et al.\ 1989, 1991; Todini \& Ferrara 2001; 
Nozawa et al.\ 2003, 2008, 2010; Bianchi \& Schneider 2007; 
Kozasa et al.\ 2009; Sarangi \& Cherchneff 2015).
However, the calculated sizes of Al$_2$O$_3$ grains are below 
$\simeq$0.03 $\mu$m, which is much smaller than the measured sizes of 
presolar Al$_2$O$_3$ grains with 0.5 $\mu$m to 1.5 $\mu$m in diameter 
(Nittler et al.\ 1998; Choi et al.\ 1998),
and such small Al$_2$O$_3$ grains are found to be almost 
completely destroyed in the hot gas within the SNRs before being ejected 
to the ISM (Nozawa et al.\ 2007; Silvia et al.\ 2010, 2012).
This seems to contradict the fact that we are observing the Al$_2$O$_3$ 
grains of a SN origin in hand on the Earth.

One of the main reasons that only small Al$_2$O$_3$ grains are produced 
in the simulations is the low number density of Al atoms in the ejecta, 
led by the relatively homogeneous ejecta of spherically symmetric SN models
(hereafter referred to as 1-D SN models).
In reality, the SN ejecta should be much more inhomogeneous and 
complicated, as is suggested from multi-dimensional hydrodynamic 
simulations 
(e.g., Kifonidis et al.\ 2003; Hammer et al.\ 2010; Joggerst et al.\ 2010) 
and various observations of SN 1987A 
(e.g., Kj\ae r et al.\ 2010; Larsson et al.\ 2013).
This implies that the formation of dust grains would proceed in the gases 
with a variety of densities, and that large grains could be formed in 
high-density clumps.

In this Letter, we investigate the formation of Al$_2$O$_3$ grains for 
wide ranges of density and cooling rate of gas, to explore the formation
condition of presolar Al$_2$O$_3$ grains as large as measured in 
meteorites. 
We show that submicron-sized Al$_2$O$_3$ grains can be produced only 
in the gas with more than ten times higher densities than those predicted 
by 1-D SN models, suggestive of the presence of dense gas clumps in the 
ejecta.
We also examine the survival of Al$_2$O$_3$ grains against the 
destruction process in SNRs.
We propose that the comparison between the calculated sizes and the 
measured sizes of presolar grains can be a novel and valuable approach 
that gives insight into the physical conditions and structure of the SN 
ejecta.

%%%%%%%%%%%%%%%%%%%%%%%%%%%%%%%%%%%%%%%%%%%%%%%%%%%%%%%%%%%%%%%%%%%%%
\section{Model of Dust Formation}
%%%%%%%%%%%%%%%%%%%%%%%%%%%%%%%%%%%%%%%%%%%%%%%%%%%%%%%%%%%%%%%%%%%%%

The formation of Al$_2$O$_3$ grains is calculated by applying the 
formula of non-steady-state dust formation in Nozawa \& Kozasa (2013).
In this formula, the formation of small clusters and the growth of grains 
are self-consistently followed under the consideration that the kinetics of 
dust formation process is controlled by collisions of key species, 
defined as the gas species that have the lowest collisional frequency
among the reactants.
The formula leads us to derive the size distribution and condensation 
efficiency of newly formed grains, given chemical reactions for the 
formation of clusters, abundances of the relevant gas species, and 
time evolutions of gas density and temperature.
The detailed prescription of the calculations of non-steady dust 
formation is given in Nozawa \& Kozasa (2013).

In the ejecta of SNe, the most likely formation site of Al$_2$O$_3$ grains 
is the O-rich layer, where Al atoms, as well as O atoms, abundantly exist.
We consider as a chemical reaction at cluster formation
2Al + 3O $\rightleftharpoons$ Al$_2$O$_3$  (Kozasa et al.\ 1989).
The initial number ratio of Al to O atoms is taken as 
$c_{\rm Al,0}/c_{\rm O,0} = 1/200$, where $c_{\rm Al,0}$ and $c_{\rm O,0}$ 
are, respectively, the number densities of Al and O atoms at a given 
initial time $t = t_0$, so that the key species are Al atoms.
This abundance ratio approximately corresponds to that in the Al-rich 
region of solar-metallicity SNe 
(see, e.g., Kozasa et al.\ 2009; Nozawa et al.\ 2010).
Note that the results of calculations are little affected by the Al/O ratio 
as long as $c_{\rm Al,0}/c_{\rm O,0} \ll 1$.

The ejecta of SNe freely expands since $\sim$1 day post-explosion,  
and the gas density is inversely proportional to the cube of time $t$.
Thus, the number density of a gas species $\tilde{c}_{i}(t)$ (where
$i$ is Al or O), without the depletion of the gas-phase atoms due to the 
formation of clusters and grains, is given by 
\begin{eqnarray}
\tilde{c}_i(t) = c_{i,0} \left( \frac{t}{t_0} \right)^{-3}.
\end{eqnarray}
As in Nozawa \& Kozasa (2013),
the gas temperature $T(t)$ is assumed to decrease as
\begin{eqnarray}
T(t) = T_0 \left( \frac{t}{t_0} \right)^{-3(\gamma - 1)},
\end{eqnarray}
where $T_0$ is the gas temperature at $t_0$, and $\gamma$ is a free 
parameter that prescribes the cooling rate.

%%%%%%%%%%%%%%%%%%%%%%%%%%%%%%%%%%%%%%%%%%%%%%%%%%%%%%%%%%%%%%%%%%%%%
\begin{deluxetable}{cccc}
\tabletypesize{\scriptsize}
\tablewidth{0pt}
\tablecaption{Numerical Constants Used for Dust Formation Calculations} 
\tablehead{ 
\colhead{$A/10^4$ K} & 
\colhead{$B$} & 
\colhead{$a_0$$^{\rm a}$} & 
\colhead{$\sigma$$^{\rm b}$} \\
& &
\colhead{(\AA)} & 
\colhead{(erg cm$^{-2}$)} 
}
\startdata
%\tableline
%\multicolumn{2}{c}{Numerical Constants Necessary for Dust Formation Calculations} \\
%\tableline
18.4788 & 
45.3542  & 
1.718  & 
690
\enddata
\footnotetext{hypothetical radius of the condensate per key molecule.}
\footnotetext{surface tension of bulk grains (Overbury et al.\ 1975).}
\end{deluxetable}
%%%%%%%%%%%%%%%%%%%%%%%%%%%%%%%%%%%%%%%%%%%%%%%%%%%%%%%%%%%%%%%%%%%%%

As the gas cools down, the formation of dust from gas can be 
realized in a supersaturated state ($S > 1$); 
$S$ is the supersaturation ratio defined as $\ln S = - \Delta g/kT$ with 
$k$ being the Boltzmann constant and $\Delta g$ the change of the 
chemical potential per key species for the formation of bulk condensate from 
the reactants, as is fomulated in Equation (30) in Nozawa \& Kozasa (2013).
Here we take $t_0$ as a time at which $S = 1$ and determine $T_0$ for 
given $c_{\rm Al,0}$ and $c_{\rm O,0}$ from the equation
\begin{eqnarray}
\ln S = -\frac{\Delta \mathring{g}}{k T_0} 
+ \ln \left( \frac{c_{\rm Al,0} k T_0}{p_{\rm s}} \right) + 
\frac{3}{2} \ln \left( \frac{c_{\rm O,0} k T_0}{p_{\rm s}} \right) = 0, ~~
\end{eqnarray}
where $\Delta \mathring{g}$ is the change of the chemical potential at the 
standrad pressure $p_{\rm s}$ and is approximated as 
$\Delta \mathring{g} / k T = -A / T + B$ with the numerical values $A$ 
and $B$ taken from Nozawa et al.\ (2003) (see Table 1).

In this study, we suppose the clusters containing more than 100 Al atoms 
as bulk grains.
This corresponds to the minimum grain radius of $a_* = 8.0$ 
\AA~(Al$_2$O$_3$ grains are assumed to be spherical).
The sticking probability of gas species is assumed to be unity for any 
sizes of clusters and grains.
In the calculations, we take $t_0 = 300$ days, so the free parameters are 
$c_{\rm Al,0}$ and $\gamma$, for which we consider the ranges of
$c_{\rm Al,0}$ = $10^4$--$10^{11}$ cm$^{-3}$ and $\gamma =$ 1.1--1.7.
The calculations are performed until the gas density becomes so low that 
grain growth is negligible.
In what follows, we mainly examine the resultant behavior of the final 
average radius $a_{\rm ave, \infty}$ and condensation efficiency 
$f_{\rm con, \infty}$ that are obtained at the end of the calculations.
The condensation efficiency is defined as the fraction of Al atoms locked 
up in Al$_2$O$_3$ grains.

%%%%%%%%%%%%%%%%%%%%%%%%%%%%%%%%%%%%%%%%%%%%%%%%%%%%%%%%%%%%%%%%%%%%%
\section{RESULTS OF DUST FORMATION CALCULATIONS}
%%%%%%%%%%%%%%%%%%%%%%%%%%%%%%%%%%%%%%%%%%%%%%%%%%%%%%%%%%%%%%%%%%%%%

%%%%%%%%%%%%%%%%%%%%%%%%%%%%%%%%%%
\begin{figure}
%\epsscale{0.8}
\epsscale{1.1}
\plotone{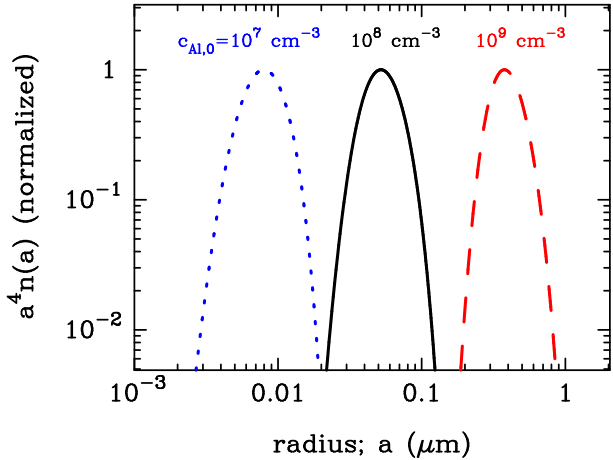}
\caption{
Final size distribution spectrum of newly formed Al$_2$O$_3$ grains 
calculated for $\gamma = 1.25$ and $t_0 = 300$ days.
The size distributions are plotted as $a^4 n(a)$ so as to represent the 
mass distribution per logarithmic grain radius, where $n(a)$ is the 
arbitrarily normalized number density of grains with radii between $a$ 
and $a + da$.
Three cases are considered for the number density of Al atoms at 
$t = t_0$; 
$c_{\rm Al,0} = 10^7$ (dotted), $10^8$ (solid), and $10^9$ cm$^{-3}$ 
(dashed).
\label{fig1}}
\end{figure}
%%%%%%%%%%%%%%%%%%%%%%%%%%%%%%%%%%

Figure 1 shows the size distributions of newly formed Al$_2$O$_3$ 
grains calculated for $\gamma = 1.25$, adopting $c_{\rm Al,0} = 10^7$, 
$10^8$, and $10^9$ cm$^{-3}$.
For these cases, all Al atoms are finally locked up in Al$_2$O$_3$ grains
(that is, $f_{\rm con, \infty} = 1$).
As seen from the figure, the size distribution is lognormal-like for any 
of $c_{\rm Al,0}$ considered here, with a narrower width for a higher 
$c_{\rm Al,0}$.
More importantly, the average radius increases with increasing 
$c_{\rm Al,0}$: 
$a_{\rm ave, \infty} = 0.0067$, 0.047, and 0.36 $\mu$m for 
$c_{\rm Al,0} = 10^7$, $10^8$, and $10^9$ cm$^{-3}$, respectively.
This is because a higher gas density leads to more efficient growth of 
grains.
The results in Figure 1 point out that, for $\gamma = 1.25$, the number 
density of Al atoms at dust formation must be higher than 
$\sim$$5 \times 10^8$ cm$^{-3}$ in order that Al$_2$O$_3$ grains with 
the radii larger than 0.25 $\mu$m ($\ga$0.5 $\mu$m in diameter) can 
be formed.

Nozawa \& Kozasa (2013) demonstrated that the formation process of 
dust grains is described in terms of the timescales of two physical quantities:
the timescale with which the supersaturation ratio $S$ increases 
$\tau_{\rm sat}$ and the collision timescale of key species $\tau_{\rm coll}$.
They found that, for C and MgSiO$_3$ grains, the average radius and 
condensation efficiency are universally scaled by one non-dimensional quantity 
$\Lambda_{\rm on} \equiv \tau_{\rm sat}(t_{\rm on})/\tau_{\rm coll}(t_{\rm on})$, 
where $t_{\rm on}$ is the onset time of dust formation ($t_{\rm on} \ga t_0)$ 
and is taken as a time at which the condensation efficiency reaches $10^{-10}$.
According to their study, it would be interesting to see if such a scaling 
relation holds for Al$_2$O$_3$ grains.

Figure 2 depicts the average grain radii $a_{\rm ave, \infty}$ and 
condensation efficiencies $f_{\rm con, \infty}$ obtained from the calculations 
with a variety of $c_{\rm Al,0}$ for each of $\gamma = 1.1$, 1.3, 1.5, and 
1.7, as a function of $\Lambda_{\rm on}$. 
In the present study, $\Lambda_{\rm on}$ is approximately written as
(Nozawa \& Kozasa 2013)
\begin{eqnarray}
\Lambda_{\rm on } 
\simeq \frac{1082}{\gamma - 1} 
\left( \frac{t_{\rm on}}{300 ~ {\rm days}} \right)
\left[ \frac{\tilde{c}_{\rm Al}(t_{\rm on})}{10^8 ~ {\rm cm}^{-3}} \right]
\left[ \frac{T(t_{\rm on})}{2000 ~ {\rm K}} \right]^{\frac{3}{2}}.
\end{eqnarray}
Figure 2 clearly shows that the results for different $\gamma$ are well 
overplotted, indicating that both $a_{\rm ave, \infty}$ and $f_{\rm con, \infty}$
are uniquely determined by $\Lambda_{\rm on}$ even for 
Al$_2$O$_3$ grains.
As is the cases of C and MgSiO$_3$ grains, the formation of Al$_2$O$_3$ 
grains can be realized at $\Lambda_{\rm on} \ga$ 1--2, and $f_{\rm con, \infty} = 1$ 
at $\Lambda_{\rm on} \ge 20$.
In addition, $a_{\rm ave, \infty}$ becomes large as $\Lambda_{\rm on}$ 
increases, which means that the final average radius is larger for a higher 
gas density and/or a slower gas cooling because $\Lambda_{\rm on}$ is 
roughly proportional to the product of gas density and cooling timescale 
that is reflected by $t_{\rm on}$. 

%%%%%%%%%%%%%%%%%%%%%%%%%%%%%%%%%%
\begin{figure}
%\epsscale{0.8}
\epsscale{1.1}
\plotone{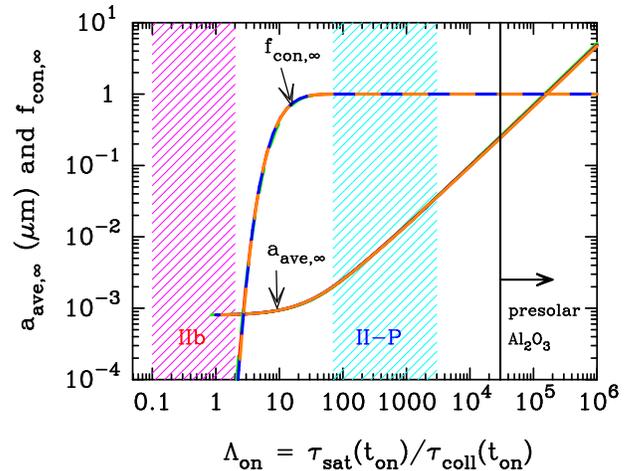}
\caption{
Dependence of the final average radii $a_{\rm ave, \infty}$ and condensation 
efficiencies $f_{\rm con, \infty}$ of newly formed Al$_2$O$_3$ grains 
on $\Lambda_{\rm on}$.
The results for four different $\gamma$ ($\gamma = 1.1$, 1.3, 1.5, 
and 1.7) are shown in each color but they are plotted as almost the 
same curve. 
The hatched regions depict the expected ranges of $\Lambda_{\rm on}$
for the formation of Al$_2$O$_3$ grains in the Al-rich region, 
referring to 1-D models of a Type II-P SN (cyan; Kozasa et al.\ 2009) 
and a Type IIb SN (Nozawa et al.\ 2010). 
The solid vertical line denotes the minimum value of $\Lambda_{\rm on}$ 
($= 3 \times 10^4$) necessary for explaining the measured sizes 
(radius of $\ge$0.25 $\mu$m) of presolar Al$_2$O$_3$ grains.
\label{fig2}}
\end{figure}
%%%%%%%%%%%%%%%%%%%%%%%%%%%%%%%%%%

In Figure 2, we also plot the ranges of $\Lambda_{\rm on}$ expected in
the ejecta of Type II-P and IIb SNe, referring to 1-D SN models used in 
Kozasa et al.\ (2009) and Nozawa et al.\ (2010), respectively.
For a Type II-P SN with the massive hydrogen envelope, the number 
density of Al atoms in the Al-rich region is estimated to be
$\tilde{c}_{\rm Al}(t_{\rm on}) \simeq$ (0.2--8) $\times 10^7$ cm$^{-3}$
at $t_{\rm on} = 300$ days.
For $\gamma \simeq 1.25$ and $T(t_{\rm on}) \simeq 2000$ K, 
this corresponds to $\Lambda_{\rm on} \simeq$ 70--3000, for which
$a_{\rm ave, \infty} \simeq$ 0.002--0.03 $\mu$m.
On the other hand, for a Type IIb SN with the small-mass envelope, 
$\tilde{c}_{\rm Al}(t_{\rm on}) \simeq$ (0.3--5) $\times 10^4$ cm$^{-3}$ 
at $t_{\rm on} = 300$ days, resulting in $\Lambda_{\rm on} \simeq$ 0.1--2.
Hence, Al$_2$O$_3$ grains are not expected to form in the Type IIb SN.
Nozawa et al.\ (2010) reported the formation of Al$_2$O$_3$ grains 
in the Type IIb SN but their average radii are less than $\simeq$8 \AA, 
which is regarded as small clusters in this study.

Our calculations show that, in order to produce Al$_2$O$_3$ grains larger 
than 0.25 $\mu$m (diameter of $\ge$0.5 $\mu$m) as measured for 
presolar grains, $\Lambda_{\rm on}$ should be higher than $3 \times 10^4$.
Given that $\gamma$ and $T(t_{\rm on})$ do not change largely,
such a high $\Lambda_{\rm on}$ could be achieved by considering the gas 
densities that are more than one order of magnitude higher than those 
presented by 1-D SN models (namely, 
$\tilde{c}_{\rm Al}(t_{\rm on}) \ga 7 \times 10^8$ cm$^{-3}$ at
$t_{\rm on} = 300$ days, see Equation (4)).
This strongly suggests that the discovered presolar Al$_2$O$_3$ grains 
were formed in dense clumps within the ejecta.
Equation (4) tells us that the formation of large Al$_2$O$_3$ grains 
may also be possible if the formation time of dust is later than 
$t_{\rm on} = 3000$ days.
However, it is too hard to keep the gas density as high as 
$\tilde{c}_{\rm Al} \simeq 10^8$ cm$^{-3}$ in such late epochs
because the gas density rapidly decreases with time (see Equation (2)).

%%%%%%%%%%%%%%%%%%%%%%%%%%%%%%%%%%%%%%%%%%%%%%%%%%%%%%%%%%%%%%%%%%%%%
\section{CONCLUSIONS AND DISCUSSION}
%%%%%%%%%%%%%%%%%%%%%%%%%%%%%%%%%%%%%%%%%%%%%%%%%%%%%%%%%%%%%%%%%%%%%

We have systematically investigated the formation of Al$_2$O$_3$ grains,
adopting a wide variety of gas densities and cooling rates. 
We show that, as in the cases of C and MgSiO$_3$ grains, the average 
radius and condensation efficiency of Al$_2$O$_3$ grains are nicely scaled 
by a non-dimensional quantity $\Lambda_{\rm on}$ defined as the ratio 
between the timescale of the supersaturation ratio and the collision timescale 
of key species at dust formation.
We also find that large Al$_2$O$_3$ grains with radii of $\ga$0.25 $\mu$m 
can be formed only in dense gas regions which have more than ten times 
higher densities than those estimated from 1-D SN models.
This points out the presence of dense clumps in the ejecta of core-collapse 
SNe.

The formation of dust in dense clumps was deduced from various 
early-phase ($\le$1000 days after explosion) observations of SN 1987A
(e.g., Lucy et al.\ 1991; Meikle et al.\ 1993; Colgan et al.\ 1994),
which can be classified as a Type II-P SN.
The recent radiative transfer models of dust emission and absoption also 
suggested the necessity of optically thick clumps to account for the 
evolution of optical to infrared emission from SN 1987A over 
$\simeq$600--9000 days
(Ercolano et al.\ 2007; Wesson et al.\ 2015; Dwek \& Arendt 2015;
Bevan \& Barlow 2015).
The density contrast between the clumps and interclumps considered by 
these works is $\simeq$10--100, which is in good agreement with the 
density enhancement needed for the formation of large Al$_2$O$_3$ 
grains (i.e., more than ten time that of 1-D SN models).
This allows us to conclude that the measured size of presolar grains is an 
independent and useful probe to constrain the clump density in the ejecta.

Our calculations show that Al$_2$O$_3$ grains cannot form in Type IIb 
SNe as a result of the too low ejecta density in the 1-D SN model.
However, the analyses of infrared emission spectra of Cas A, which was 
identified as Type IIb through the light echo (Krause et al.\ 2008),
have suggested the presense of newly formed Al$_2$O$_3$ grains 
(Douvion et al.\ 2001; Rho et al.\ 2008).
This contradiction can also be resolved by considering the dense gas 
clumps in the ejecta;
from Figure 2, we can estimate that the density of gas clumps required for 
the formation of Al$_2$O$_3$ grains (that is, $\Lambda_{\rm on} \ga 2$) 
is more than ten times that of the 1-D Type IIb SN model.
We also note that the recent observation of near-infrared extinction implies 
the existence of large ($\ga$0.1 $\mu$m) Si grains in Cas A, providing 
another indication of dust formation in dense clumps (Lee et al.\ 2015).

Next we discuss the survival of Al$_2$O$_3$ grains formed in the ejecta 
against the destruction by SN shocks on the basis of a dust evolution 
model in SNRs (Nozawa et al.\ 2007).
This model assumes a spherically symmetry shock but it has been 
shown that the destruction efficiency of dust is not greatly different 
from those by multi-dimensional hydrodynamic simulations 
(Silvia et al.\ 2010, 2012).
As the initial condition of the ejecta, we here consider the core-envelope 
structure of $3+12$ $M_\odot$ (implicitly assuming Type II-P SNe).
The other input parameters are the explosion energy and the gas 
density in the ISM, for which we take reasonable values of 
$1.8 \times 10^{51}$ erg and 1.0 cm$^{-3}$, respectively.

%%%%%%%%%%%%%%%%%%%%%%%%%%%%%%%%%%
\begin{figure}
%\epsscale{0.8}
\epsscale{1.1}
\plotone{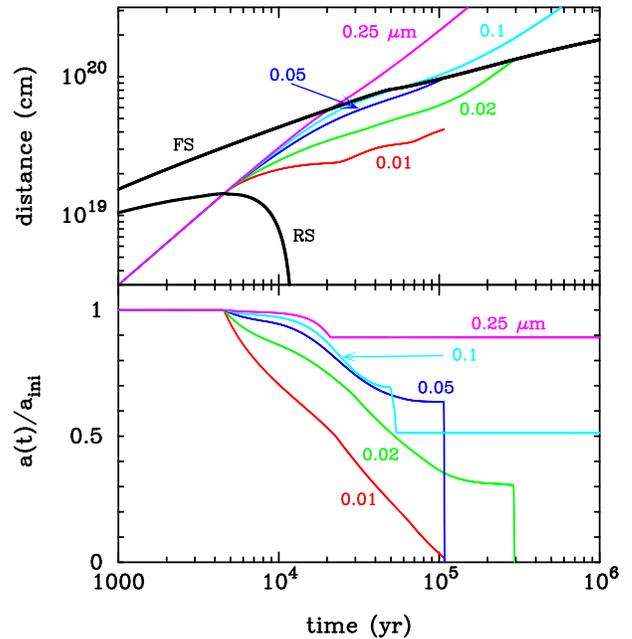}
\caption{
Time evolutions of the positions (upper) and the ratios of radii to the 
initial ones (lower) of Al$_2$O$_3$ grains in a SNR.
The results are plotted for the initial grain radii of 0.01, 0.02, 0.05, 0.1, 
and 0.25 $\mu$m.
In the upper panel, the thick solid curves depict the trajectories of the 
forward shock (FS) and reverse shock (RS).
\label{fig3}}
\end{figure}
%%%%%%%%%%%%%%%%%%%%%%%%%%%%%%%%%%

The result of the calculation is given in Figure 3, which shows that
small grains with the initial radii of $a_{\rm ini} \la 0.01$ $\mu$m are 
completely destroyed in the shocked gas.
Grains with 0.02 $\mu$m $\la a_{\rm ini} < 0.1$ $\mu$m are eroded
in the hot gas and are finally destroyed as soon as they encounter the 
cool dense shell that is formed behind the forward shock after 
$\simeq 5 \times 10^4$ years.
On the other hand, Al$_2$O$_3$ grains with $a_{\rm ini} \ga$ 
0.2 $\mu$m can be ejected from SNe without reducing their size 
significantly.
This simple calculation illustrates that, once large Al$_2$O$_3$ grains 
are produced in dense clumps, they are likely to survive the destruction 
in SNRs and to be easily transported to the 
ISM.\footnote{In this calculation, we do not include the destruction of 
dust by non-thermal sputtering resulting from the high-velocity motion 
of dust after being injected into the ISM, in order to focus on the 
survivability of dust in the SNR.}
Thus, we suggest that the formation of submicron-sized Al$_2$O$_3$ 
grains in dense clumps is also indispensable in order that newly formed 
Al$_2$O$_3$ grains can endure the destruction by shocks and be 
incorporated into nearby molecular clouds and protoplanetary disks.

The identification of Al$_2$O$_3$ grains as a SN origin comes from 
their relatively high $^{18}$O/$^{16}$O ratio 
(so-called Group 4 grains, Nittler et al.\ 1997; Choi et al.\ 1998) 
or large enrichment of $^{16}$O (represented by Grain T84,
Nittler et al.\ 1998).
Since we consider the formation of Al$_2$O$_3$ grains in the O-rich 
layer where $^{16}$O is rich, they may be categorized as T84-like
grains (or Group 3 grains with the moderate enhancement of $^{16}$O).
On the other hand, to explain the $^{18}$O-enriched composition of 
Group 4 grains, the extensive mixing between different layers in the 
ejecta is invoked with the relatively $^{18}$O-rich hydrogen envelope 
being the dominant component (Nittler et al.\ 2008).
However, even if the large-scale mixing takes place, the molecular 
diffusion lengths are much smaller than the typical size of gas clumps 
so that the microscopic mixing of elements may be very ineffective
(Deneault et al.\ 2003).
Therefore, the origin of oxygen isotopic composition of the Group 4 
Al$_2$O$_3$ grains, as well as their formation process, is still a 
challenging problem.

We conclude that dense gas clumps are necessary for the formation of
submicron-sized Al$_2$O$_3$ grains as discovered in meteorites.
The presence of such dense clumps in the O-rich layer may also cause 
silicate grains to be formed with very large radii, compared to those based 
on 1-D SN models.
Given that the radii of silicate grains are generally larger than those of 
Al$_2$O$_3$ grains by a factor of about ten (e.g., Nozawa et al.\ 2003),
the discovery of $^{16}$O-enriched micron-sized ($\simeq$1--10 $\mu$m) 
silicate grains in meteorites will serve as further evidence for dense clumps 
in the SN ejecta.

%%%%%%%%%%%%%%%%%%%%%%%%%%%%%%%%%%%%%%%%%%%%%%%%%%%%%%%%%%%%%%%%%%%%%
\acknowledgments

We thank Masaomi Tanaka for useful comments.
We are grateful to the anonymous referee for critical comments
that improved the manuscript.
This work has been supported in part by the JSPS Grant-in-Aid for Scientific 
Research (23224004, 26400223).
%%%%%%%%%%%%%%%%%%%%%%%%%%%%%%%%%%%%%%%%%%%%%%%%%%%%%%%%%%%%%%%%%%%%%

\end{document}